\documentclass{article}
\usepackage{amsfonts}
\usepackage{amsmath}
\usepackage{amssymb}
\usepackage{graphicx}

\begin{document}

\title{A two-parameter generalization of the complete elliptic integral of
second kind}
\author{Victor B\^{a}rsan  \\
Department of Theoretical Physics, NIPNE,\\
Str. Atomistilor no. 407, Bucharest-Magurele, Romania}
\maketitle

\begin{abstract}
A two-parameter generalization of the complete elliptic integral of second
kind is expressed in terms of the Appell function $F_{4}$. This function is
further reduced to a quite simple bilinear form in the complete elliptic
integrals $K$ and $E$. The physical applications are briefly mentioned.
\end{abstract}

\bigskip The integral

\begin{equation}
K\left( k_{1},k_{2}\right) =\int_{0}^{\pi/2}\int_{0}^{\pi/2}\frac{d\theta
d\varphi}{\sqrt{1-k_{1}^{2}\sin^{2}\theta-k_{2}^{2}\sin^{2}\varphi}} 
\label{1}
\end{equation}
can be considered a two-parameter generalization of the complete elliptic
integral of first kind $K.$ Its dependence of the parameters $k_{1},k_{2}$
is (\cite{BY}, 531.07; \cite{PR}, 3.1.5.16):%
\begin{equation}
K\left( k_{1},k_{2}\right) =\frac{2}{1+k_{2}^{\prime}}K\left( k_{3}\right)
K^{\prime}\left( k_{4}\right)   \label{2}
\end{equation}
where:

\begin{equation}
k_{3}=\frac{k_{1}^{\prime}-\sqrt{1-k_{1}^{2}-k_{2}^{2}}}{1+k_{2}^{\prime}},%
\text{ }k_{4}=\frac{k_{1}^{\prime}+\sqrt{1-k_{1}^{2}-k_{2}^{2}}}{%
1+k_{2}^{\prime}},\text{ }k_{1}^{2}+k_{2}^{2}<1   \label{3}
\end{equation}

However, the integral

\begin{equation}
E\left( k_{1},k_{2}\right) =\int_{0}^{\pi/2}\int_{0}^{\pi/2}\sqrt {%
1-k_{1}^{2}\sin^{2}\theta-k_{2}^{2}\sin^{2}\varphi}d\theta d\varphi 
\label{4}
\end{equation}
which can be considered a two-parameter generalization of the complete
elliptic integral of second kind, has not a compact form, at the best of
author's knowledge. This article aims to derive such a result for $E\left(
k_{1},k_{2}\right) ,$ a function which is interesting for applications in
statistical physics: it is proportional to the statistical sum of a three
dimensional system of weakly coupled Ginzburg-Landau chains (for the two
dimensional variant of the problem, see (\cite{VB}).

The double integral (\ref{1}) can be re-written as:

\begin{equation}
E\left( k_{1},k_{2}\right) =\frac{1}{4}\sqrt{1-\frac{k_{1}^{2}}{2}-\frac{%
k_{2}^{2}}{2}}\int_{0}^{\pi}\int_{0}^{\pi}\sqrt{1+A\cos x+B\cos y}dxdy 
\label{5}
\end{equation}
with

\begin{equation}
A=\frac{\frac{k_{1}^{2}}{2}}{1-\frac{k_{1}^{2}}{2}-\frac{k_{2}^{2}}{2}},%
\text{ }B=\frac{\frac{k_{2}^{2}}{2}}{1-\frac{k_{1}^{2}}{2}-\frac{k_{2}^{2}}{2%
}}   \label{6}
\end{equation}

Using the formula (\cite{Grad} 2.617.5, 3.671.1):

\begin{equation}
\int_{0}^{\pi}\sqrt{a+b\cos x}dx=2\sqrt{a+b}E\left( k\right)   \label{7}
\end{equation}
where

\begin{equation}
k^{2}=\frac{2b}{a+b};\quad a,b>0   \label{8}
\end{equation}
we get:

\begin{equation}
\int_{0}^{\pi}\sqrt{1+A\cos x+B\cos y}dx=2\sqrt{1+A+B\cos y}E\left( \sqrt{%
\frac{2A}{1+A+B\cos y}}\right)   \label{9}
\end{equation}

Let us put

\begin{equation}
I=\int_{0}^{\pi}\int_{0}^{\pi}\sqrt{1+A\cos x+B\cos y}dxdy   \label{10}
\end{equation}
We have:

\begin{equation}
I=2\int_{0}^{\pi}\sqrt{1+A+B\cos y}E\left( \sqrt{\frac{2A}{1+A+B\cos y}}%
\right) dy   \label{11}
\end{equation}

Changing the variable:

\begin{equation}
z^{2}=\frac{2A}{1+A+B\cos y}   \label{12}
\end{equation}
we get:

\begin{equation}
I=\sqrt{2A}z_{0}z_{\pi}\int_{z_{0}}^{z_{\pi}}\frac{E\left( z\right) dz}{z^{2}%
\sqrt{\left( z^{2}-z_{0}^{2}\right) \left( z_{\pi}^{2}-z^{2}\right) }} 
\label{13}
\end{equation}
with

\begin{equation}
z_{0}^{2}=\frac{2A}{1+A+B};\quad z_{\pi}^{2}=\frac{2A}{1+A-B}   \label{14}
\end{equation}

Writing the elliptic integral $E$ as a hypergeometric series:

\begin{equation}
E\left( z\right) =\frac{\pi}{2}F\left( -\frac{1}{2},\frac{1}{2}%
;1;z^{2}\right) =\frac{\pi}{2}\sum_{n=0}^{\infty}\frac{\left( -\frac{1}{2}%
\right) _{n}\left( \frac{1}{2}\right) _{n}}{n!\left( 1\right) _{n}}z^{2n} 
\label{15}
\end{equation}
the integral (\ref{13}) becomes:

\begin{equation}
I=2\pi\sqrt{2A}z_{0}z_{\pi}\sum_{n=0}^{\infty}\frac{\left( -\frac{1}{2}%
\right) _{n}\left( \frac{1}{2}\right) _{n}}{\left( n!\right) ^{2}}%
\int_{z_{0}}^{z_{\pi}}\frac{z^{2n-2}dz}{\sqrt{\left( z^{2}-z_{0}^{2}\right)
\left( z_{\pi}^{2}-z^{2}\right) }}   \label{16}
\end{equation}

Adapting the formula (\cite{BY}, 218.15)

\begin{equation}
\int_{y}^{a}\frac{R\left( t^{2}\right) dt}{\sqrt{\left( a^{2}-t^{2}\right)
\left( t^{2}-b^{2}\right) }}=g\int_{0}^{u_{1}}R\left( a^{2}\mathtt{dn}%
^{2}u\right) du   \label{17}
\end{equation}
with the conditions

\begin{equation}
a>y\geq b>0   \label{18}
\end{equation}
and the notations:

\begin{equation}
k^{2}=\frac{a^{2}-b^{2}}{a^{2}};\quad g=\frac{1}{a};\quad\sin\psi=\sqrt {%
\frac{a^{2}-y^{2}}{a^{2}-b^{2}}}   \label{19}
\end{equation}
to the integral in (\ref{17}), we must put:

\begin{equation}
a=z_{\pi},\quad y=b=z_{0}   \label{20}
\end{equation}

The values of parameters in (\ref{17}) are:

\begin{equation}
k^{2}=\frac{2B}{1+A+B},\quad g=\sqrt{\frac{1+A-B}{2A}},\quad\sin \psi=1 
\label{21}
\end{equation}
so the integral is complete and takes the form:

\begin{equation}
\int_{z_{0}}^{z_{\pi}}\frac{R\left( z^{2}\right) dz}{\sqrt{\left(
z^{2}-z_{0}^{2}\right) \left( z_{\pi}^{2}-z^{2}\right) }}=\sqrt {\frac{1+A-B%
}{2A}}\int_{0}^{K}R\left( \frac{2A}{1+A-B}\mathtt{dn}^{2}u\right) du 
\label{22}
\end{equation}

But (\cite{BY} 810.00)

\begin{equation}
P_{m-1/2}\left( x\right) =\frac{2}{\pi}\left( x+\sqrt{x^{2}-1}\right)
^{m-1/2}\int_{0}^{K}\mathtt{dn}^{2m}udu   \label{23}
\end{equation}
with

\begin{equation}
k^{2}=2\left[ x\sqrt{x^{2}-1}+1-x^{2}\right]   \label{24}
\end{equation}
Inverting (\ref{24}), we obtain:

\begin{equation}
x^{2}=\frac{\left( 1-\frac{k^{2}}{2}\right) ^{2}}{1-k^{2}}   \label{25}
\end{equation}
or, with (\ref{21}) and (\ref{6}),

\begin{equation}
x^{2}=\frac{\left( 1+A\right) ^{2}}{\left( 1+A\right) ^{2}-B^{2}} 
\label{26}
\end{equation}
and (\ref{23}) gives:

\begin{equation}
\int_{0}^{K}\mathtt{dn}^{2m}udu=\frac{\pi}{2}\left( \frac{1+A-B}{1+A+B}%
\right) ^{\frac{1}{2}\left( m-\frac{1}{2}\right) }P_{m-\frac{1}{2}}\left( 
\frac{1+A}{\sqrt{\left( 1+A\right) ^{2}-B^{2}}}\right)   \label{27}
\end{equation}

So, finally,

\begin{equation}
\int_{z_{0}}^{z_{\pi}}\frac{z^{2n-2}dz}{\sqrt{\left( z^{2}-z_{0}^{2}\right)
\left( z_{\pi}^{2}-z^{2}\right) }}=\frac{\pi}{2}\left( \frac{2A}{\sqrt{%
\left( 1+A\right) ^{2}-B^{2}}}\right) ^{n-\frac{3}{2}}P_{n-\frac {3}{2}%
}\left( \frac{1+A}{\sqrt{\left( 1+A\right) ^{2}-B^{2}}}\right)   \label{28}
\end{equation}
and 
\begin{equation}
I=\pi^{2}\sqrt{2A}z_{0}z_{\pi}\sum_{n=0}^{\infty}\frac{\left( -\frac{1}{2}%
\right) _{n}\left( \frac{1}{2}\right) _{n}}{\left( n!\right) ^{2}}t^{n-\frac{%
3}{2}}P_{n-\frac{3}{2}}\left( x\right)   \label{29}
\end{equation}
with the notations:

\begin{equation}
t=\frac{2A}{\sqrt{\left( 1+A\right) ^{2}-B^{2}}},\quad x=\frac{1+A}{\sqrt{%
\left( 1+A\right) ^{2}-B^{2}}}   \label{30}
\end{equation}
Observing that:

\begin{equation*}
z_{0}z_{\pi}=t 
\end{equation*}
eq. (28) can be written as:

\begin{equation}
I=\pi^{2}\sqrt{2A}\sum_{n=0}^{\infty}\frac{\left( -\frac{1}{2}\right)
_{n}\left( \frac{1}{2}\right) _{n}}{n!\left( 1\right) _{n}}t^{n-\frac {1}{2}%
}P_{n-\frac{3}{2}}\left( x\right)   \label{31}
\end{equation}

Writing

\begin{equation*}
\left( \frac{1}{2}\right) _{k}=-2\left( k-\frac{1}{2}\right) \left( -\frac{1%
}{2}\right) _{k} 
\end{equation*}
the previous expression takes the form:

\begin{equation}
I=-2\pi^{2}\sqrt{2A}\sum_{n=0}^{\infty}\left[ \frac{\left( -\frac{1}{2}%
\right) _{n}}{n!}\right] ^{2}\left( n-\frac{1}{2}\right) t^{n-\frac {1}{2}%
}P_{n-\frac{3}{2}}\left( x\right)   \label{32}
\end{equation}

The sum (\ref{29}) can be put in a compact form using the formula (\cite{PR2}
6.5.1.18 for $\mu=0$)

\begin{equation}
\sum_{k=0}^{\infty}\frac{\left( 1+\nu\right) _{k}\left( 1+\nu\right) _{k}}{%
k!\left( 1\right) _{k}}t^{k}P_{k+\nu}\left( x\right) =\left( \frac {x+1}{2}%
\right) ^{-\nu-1}F_{4}\left( 1+\nu,1+\nu;1,1;\frac{2t}{1+x},\frac{x-1}{x+1}%
\right)   \label{33}
\end{equation}
With $\nu=-\frac{3}{2}:$

\begin{equation}
\sum_{k=0}^{\infty}\frac{\left( \left( -\frac{1}{2}\right) _{k}\right) ^{2}}{%
\left( k!\right) ^{2}}t^{k}P_{k-3/2}\left( x\right) =\left( \frac{x+1}{2}%
\right) ^{1/2}F_{4}\left( -\frac{1}{2},-\frac{1}{2};1,1;\frac{2t}{1+x},\frac{%
x-1}{x+1}\right)   \label{34}
\end{equation}

\begin{equation}
I=-2\pi^{2}\sqrt{2A}\left( \frac{x+1}{2}\right) ^{1/2}t\frac{\partial }{%
\partial t}t^{-1/2}F_{4}\left( -\frac{1}{2},-\frac{1}{2};1,1;\frac{2t}{1+x},%
\frac{x-1}{x+1}\right)   \label{35}
\end{equation}
where the function $\frac{\partial}{\partial t}F_{4}$ must be taken in the
point $\left( t,x\right) $ defined by (\ref{30}). The arguments of the
Appell function $F_{4}$ in (\ref{34}) can be expressed in terms of the
parameters $k_{1},k_{2}$ using the relations:

\begin{equation}
t=\frac{k_{1}^{2}}{k_{2}^{\prime}},\quad x=\frac{1}{2}\frac{1+k_{2}^{\prime2}%
}{k_{2}^{\prime}}   \label{36}
\end{equation}

\begin{equation*}
\frac{2t}{1+x}=\frac{4k_{1}^{2}}{\left( 1+k_{2}^{\prime}\right) ^{2}};\ \ \ 
\frac{x-1}{x+1}=\frac{\left( 1-k_{2}^{\prime}\right) ^{2}}{\left(
1+k_{2}^{\prime}\right) ^{2}} 
\end{equation*}

The parameters of the Appell function are very particular, and the function
can be "reduced", using the formula (\cite{Slater}, 8.4.14):

\begin{equation}
F_{4}\left[ a,b;c,c^{\prime};x\left( 1-y\right) ,y\left( 1-x\right) \right]
=\sum_{r=0}^{\infty}\left( a\right) _{r}\left( b\right) _{r}\frac{\left(
1+a+b-c-c^{\prime}\right) }{r!\left( c\right) \left( c^{\prime}\right) }%
x^{r}y^{r}\cdot   \label{37}
\end{equation}

\begin{equation*}
\cdot F\left( a+r,b+r;c+r;x\right) F\left( a+r,b+r;c^{\prime}+r;y\right) 
\end{equation*}
with $F$ - the Gaussian hypergeometric function.

In our case,

\begin{equation}
a=b=-\frac{1}{2},c=c^{\prime}=1   \label{38}
\end{equation}
so:

\begin{equation*}
F_{4}\left[ -\frac{1}{2},-\frac{1}{2};1,1;u\left( 1-v\right) ,v\left(
1-u\right) \right] =F\left( -\frac{1}{2},-\frac{1}{2};1;u\right) \cdot
F\left( -\frac{1}{2},-\frac{1}{2};1;v\right) - 
\end{equation*}

\begin{equation}
-\frac{1}{2}uv\left( \frac{1}{2},\frac{1}{2};2;u\right) \cdot F\left( \frac{1%
}{2},\frac{1}{2};2;v\right) +\frac{1}{16}u^{2}v^{2}\cdot F\left( \frac{3}{2},%
\frac{3}{2};3;u\right) \cdot F\left( \frac{3}{2},\frac{3}{2};3;v\right) 
\label{39}
\end{equation}

In order to apply (\ref{37}), we have to solve the system:

\begin{equation}
u\left( 1-v\right) =\frac{2t}{x+1},\qquad v(1-u)=\frac{x-1}{x+1}   \label{40}
\end{equation}
It has the solution:

\begin{equation}
v=\frac{x-t\pm\sqrt{1-2tx+t^{2}}}{x+1},\qquad u=\frac{1+t\pm\sqrt{1-2tx+t^{2}%
}}{x+1}   \label{41}
\end{equation}
or

\begin{equation}
u=2\frac{k_{1}^{2}+k_{2}^{\prime}\pm k_{1}^{\prime}\sqrt{%
1-k_{1}^{2}-k_{2}^{2}}}{\left( 1+k_{2}^{\prime}\right) ^{2}};\ \ \ \ v=\frac{%
1-2k_{1}^{2}+k_{2}^{\prime2}\pm2k_{1}^{\prime}\sqrt{1-k_{1}^{2}-k_{2}^{2}}}{%
\left( 1+k_{2}^{\prime}\right) ^{2}}   \label{42}
\end{equation}

\begin{equation}
\sqrt{1-2tx+t^{2}}=\sqrt{1-k_{1}^{2}-k_{2}^{2}}\frac{k_{1}^{\prime}}{%
k_{2}^{\prime}}=k^{\prime}\frac{k_{1}^{\prime}}{k_{2}^{\prime}}   \label{43}
\end{equation}

Also, using the formula:

\begin{equation}
\sqrt{A\pm\sqrt{B}}=\sqrt{\frac{A+m}{2}}\pm\sqrt{\frac{A-m}{2}}   \label{44}
\end{equation}
we get:

\begin{equation}
\sqrt{u}=\frac{1}{1+k_{2}^{\prime}}\left( \sqrt{\left( 1+k_{1}\right) \left(
k_{2}^{\prime}+k_{1}\right) }\pm\sqrt{\left( 1-k_{1}\right) \left(
k_{2}^{\prime}-k_{1}\right) }\right)   \label{45}
\end{equation}

\begin{equation}
\sqrt{v}=\frac{1}{1+k_{2}^{\prime}}\left( k_{1}^{\prime}\pm\sqrt{%
1-k_{1}^{2}-k_{2}^{2}}\right)   \label{46}
\end{equation}

If we put:

\begin{equation}
k^{2}=k_{1}^{2}+k_{2}^{2};\ \ \ \
k^{\prime2}=1-k^{2}=1-1-k_{1}^{2}-k_{2}^{2}   \label{47}
\end{equation}
we can write $u,v$ as:

\begin{equation}
u=2\frac{k_{1}^{2}+k_{2}^{\prime}\pm k_{1}^{\prime}k^{\prime}}{\left(
1+k_{2}^{\prime}\right) ^{2}};\ \ \ \ v=\frac{1-2k_{1}^{2}+k_{2}^{\prime2}%
\pm2k_{1}^{\prime}k^{\prime}}{\left( 1+k_{2}^{\prime}\right) ^{2}} 
\label{48}
\end{equation}

Also,

\begin{equation}
\frac{\partial}{\partial t}u=\dot{u}=\frac{1}{\left( 1+k_{2}^{\prime}\right)
^{2}}\frac{k_{2}^{\prime}}{k^{\prime}k_{1}^{\prime}}\left[
2k^{\prime}k_{1}^{\prime}\mp\left( 2k^{\prime}+k_{2}^{2}\right) \right] 
\label{49}
\end{equation}

\begin{equation}
\frac{\partial}{\partial t}v=\dot{v}=\frac{1}{\left( 1+k_{2}^{\prime}\right)
^{2}}\frac{k_{2}^{\prime}}{k^{\prime}k_{1}^{\prime}}\left[
-2k^{\prime}k_{1}^{\prime}\mp\left( 2k^{\prime}+k_{2}^{2}\right) \right] 
\label{50}
\end{equation}

\bigskip

Both roots ($\pm$) satisfy the relation:

\begin{equation}
0<u,v<1   \label{51}
\end{equation}
which results, in fact, from (\ref{40}). By direct substitution in the
formula (\ref{39}), for $k_{1},\ k_{2}\ \rightarrow0,$ we can check that
only the roots $u_{-},\ v_{-}$ must be used and finally we obtain:

\begin{equation*}
F_{4}\left( -\frac{1}{2},-\frac{1}{2};1,1;\frac{2t}{1+x},\frac{x-1}{x+1}%
\right) = 
\end{equation*}

\begin{equation}
=F\left( -\frac{1}{2},-\frac{1}{2};1;u\right) F\left( -\frac{1}{2},-\frac{1}{%
2};1;v\right) -\frac{1}{2}uvF\left( \frac{1}{2},\frac{1}{2};2;u\right)
F\left( \frac{1}{2},\frac{1}{2};2;v\right) +   \label{52}
\end{equation}

\begin{equation*}
+\frac{1}{16}u^{2}v^{2}F\left( \frac{3}{2},\frac{3}{2};3;u\right) F\left( 
\frac{3}{2},\frac{3}{2};3;v\right) 
\end{equation*}
where

\begin{equation*}
u=u(x,t)=u\left( k_{1},k_{2}\right) =u_{-},\qquad v=v(x,t)=v\left(
k_{1},k_{2}\right) =v_{-} 
\end{equation*}
are defined by (\ref{41}, \ref{42}).

Let us express the r.h.s. of the formula (\ref{52}) in terms of elliptic
integrals. According to Slater (1.4.1.1):

\begin{equation}
\frac{d}{dz}F\left( a,b;c;z\right) =\frac{ab}{c}F\left( a+1,b+1;c+1;z\right) 
\label{53}
\end{equation}
Consequently,

\begin{equation}
F\left( \frac{1}{2},\frac{1}{2};2;z\right) =4\frac{d}{dz}F\left( -\frac {1}{2%
},-\frac{1}{2};1;z\right) =4F^{\prime}\left( -\frac{1}{2},-\frac{1}{2}%
;1;z\right)   \label{54}
\end{equation}

\begin{equation}
F\left( \frac{3}{2},\frac{3}{2};3;z\right) =32\frac{d^{2}}{dz^{2}}F\left( -%
\frac{1}{2},-\frac{1}{2};1;z\right) =32F^{\prime\prime}\left( -\frac{1}{2},-%
\frac{1}{2};1;z\right)   \label{55}
\end{equation}
and

\begin{equation}
F_{4}\left( -\frac{1}{2},-\frac{1}{2};1,1;\frac{2t}{1+x},\frac{x-1}{x+1}%
\right) =   \label{56}
\end{equation}

\begin{align*}
& =F\left( -\frac{1}{2},-\frac{1}{2};1;u\right) F\left( -\frac{1}{2},-\frac{1%
}{2};1;v\right) -8uvF^{\prime}\left( -\frac{1}{2},-\frac{1}{2};1;u\right)
F^{\prime}\left( -\frac{1}{2},-\frac{1}{2};1;v\right) + \\
& +64u^{2}v^{2}F^{\prime\prime}\left( -\frac{1}{2},-\frac{1}{2};1;u\right)
F^{\prime\prime}\left( -\frac{1}{2},-\frac{1}{2};1;v\right)
\end{align*}

It is easy to show that:

\begin{equation}
F\left( -\frac{1}{2},-\frac{1}{2};1;z\right) =4z^{1/2}\left( 1-z\right) ^{2}%
\frac{d}{dz}z\frac{d}{dz}z^{1/2}F\left( \frac{1}{2},\frac{1}{2};1;z\right) 
\label{57}
\end{equation}
Also,

\begin{equation}
F\left( \frac{1}{2},\frac{1}{2};1;z\right) =\frac{2}{\pi}K\left( \sqrt {z}%
\right)   \label{58}
\end{equation}
With the formulae of the derivatives of the complete elliptic integrals (see
for instance (\cite{BY} 710.00, 710.02)), we find:

\begin{equation}
F\left( -\frac{1}{2},-\frac{1}{2};1;z\right) =\frac{2}{\pi}\left[ 2E\left( 
\sqrt{z}\right) -\left( 1-z\right) K\left( \sqrt{z}\right) \right] 
\label{59}
\end{equation}
Similarly,

\begin{equation}
F^{\prime}\left( -\frac{1}{2},-\frac{1}{2};1;z\right) =\frac{1}{\pi z}\left[
E\left( \sqrt{z}\right) -\left( 1-z\right) K\left( \sqrt {z}\right) \right] 
\label{60}
\end{equation}

\begin{equation}
F^{\prime\prime}\left( -\frac{1}{2},-\frac{1}{2};1;z\right) =\frac{1}{2\pi
z^{2}}\left[ \left( 2-z\right) K\left( \sqrt{z}\right) -2E\left( \sqrt{z}%
\right) \right]   \label{61}
\end{equation}

Finally, we obtain:

\begin{equation}
F_{4}\left( -\frac{1}{2},-\frac{1}{2};1,1;\frac{2t}{1+x},\frac{x-1}{x+1}%
\right) =   \label{62}
\end{equation}

\begin{equation*}
=\frac{4}{\pi^{2}}\left\{ 66E\left( \sqrt{u}\right) E\left( \sqrt {v}\right)
-32\left( 2-v\right) E\left( \sqrt{u}\right) K\left( \sqrt {v}\right)
-32\left( 2-u\right) E\left( \sqrt{v}\right) K\left( \sqrt {u}\right)
\right. 
\end{equation*}%
\begin{equation*}
\left. +\left[ 63-31\left( u+v\right) +15uv\right] K\left( \sqrt {u}\right)
K\left( \sqrt{v}\right) \right\} . 
\end{equation*}

Also,

\begin{equation}
\frac{\partial}{\partial t}E\left( \sqrt{u}\right) =\frac{1}{2u}\left[
E\left( \sqrt{u}\right) -K\left( \sqrt{u}\right) \right] \dot {u} 
\label{63}
\end{equation}

\begin{equation}
\frac{\partial}{\partial t}K\left( \sqrt{u}\right) =\frac{1}{2u}\left[ \frac{%
E\left( \sqrt{u}\right) }{1-u}-K\left( \sqrt{u}\right) \right] \dot{u} 
\label{64}
\end{equation}

In spite of the reducibility of the Appell function $F_{4}$, the most
compact formula for $E\left( k_{1},k_{2}\right) $ is:

\begin{equation}
E\left( k_{1},k_{2}\right) =-\frac{\pi^{2}}{2}k_{1}\left( \frac{x+1}{2}%
\right) ^{1/2}t\frac{\partial}{\partial t}\left[ t^{-1/2}F_{4}\left( -\frac{1%
}{2},-\frac{1}{2};1,1;\frac{2t}{1+x},\frac{x-1}{x+1}\right) \right] 
\label{65}
\end{equation}
where the r.h.s. must be taken in the point $\left( t,x\right) $ defined by (%
\ref{30})

For the applications in statistical mechanics, the particular case $%
k_{1}=k_{2}$ is important. The arguments of the Appell function simplifies,
in the sense that:

\begin{equation}
F_{4}\left( -\frac{1}{2},-\frac{1}{2};1,1;\frac{2t}{1+x},\frac{x-1}{x+1}%
\right) =F_{4}\left( -\frac{1}{2},-\frac{1}{2};1,1;4\xi,\xi^{2}\right) 
\label{66}
\end{equation}
with

\begin{equation}
\xi=\frac{1+A-\sqrt{1+2A}}{A}   \label{67}
\end{equation}
The convergence of the double series defined by the Appell functions is
assured if (\cite{BA})

\begin{equation}
2\sqrt{\xi}+\xi<1   \label{68}
\end{equation}
or

\begin{equation}
A<\frac{1}{2}.   \label{69}
\end{equation}
In terms of the parameters $k_{1},k_{2},$

\begin{equation}
k_{1}=k_{2}<\frac{1}{\sqrt{2}}   \label{70}
\end{equation}

The formulae (\ref{65}), (\ref{66}) allow us to find the exact
thermodynamics of a three dimensional anisotropic physical system, namely a
regular array of parallel Ginzburg-Landau chains.

\bigskip The author is indebted to Prof. D. Grecu and Prof. N. Grama for
useful discussions and suggestions. 

\end{document}